\documentclass[review]{elsarticle}
\usepackage{lineno}
\usepackage[hidelinks]{hyperref}

\usepackage{multirow,setspace,amssymb,amsmath,graphicx,color,rotating,subfigure,url}
\usepackage{natbib}
\setcitestyle{authoryear,longnamesfirst,open={(},close={)}} 
\setcitestyle{citepsep={;}}
\usepackage{textcomp}
\usepackage{CJK}
\usepackage{bm}%
\usepackage{rotating}
\usepackage{epsfig}% Include figure files
\usepackage{dcolumn}% Align table columns on decimal point
\usepackage{epstopdf}

\usepackage{graphicx}% Include figure files
\usepackage{dcolumn}% Align table columns on decimal point
\usepackage{bm}% bold math
\usepackage{color}% bold math
\usepackage{amsmath}
\usepackage[utf8]{inputenc}
\usepackage{doi}
\usepackage[T1]{fontenc}
\usepackage{diagbox}
\usepackage{amssymb}
\usepackage{enumerate}
\usepackage[ruled,linesnumbered]{algorithm2e}
\usepackage{amsmath}
\usepackage{mathrsfs}
\usepackage{booktabs}
\usepackage[ruled]{algorithm2e}

\usepackage{indentfirst}
\usepackage[a4paper, margin=1in]{geometry}

\begin{document}
\begin{frontmatter}

\title{Structural-Aware Key Node Identification in Hypergraphs via Representation Learning and Fine-Tuning}

% Group authors per affiliation:
\author[inst1]{Xiaonan Ni}
\author[inst1]{Guangyuan Mei}
\author[inst1]{Su-Su Zhang}
\author[inst4]{Yang Chen} 
%% Corresponding author
%\corref{ca1}}\ead{chenyang@fudan.edu.cn}
\author[inst3]{Xin Xu\corref{ca1}}\ead{XinXu@lixin.edu.cn}
\author[inst1]{Chuang Liu \corref{ca1}}\ead{liuchuang@hznu.edu.cn}
 %% Corresponding author

\author[inst1,inst2]{Xiu-Xiu Zhan\corref{ca1}}\ead{zhanxiuxiu@hznu.edu.cn}  %% Corresponding author

\cortext[ca1]{Corresponding authors.}

\address[inst1]{Research Center for Complexity Sciences, Hangzhou Normal University, Hangzhou, 311121, P. R. China}
\address[inst2]{College of Media and International Culture, Zhejiang University, Hangzhou 310058, P. R. China}
\address[inst4]{Shanghai Key Lab of Intelligent Information Processing, School of Computer Science, Fudan University, Shanghai 200433, P. R. China}
\address[inst3]{School of Information Management, Shanghai Lixin University of Accounting and Finance, Shanghai 201209, P. R. china}
\begin{abstract}

Evaluating node importance is a critical aspect of analyzing complex systems, with broad applications in digital marketing, rumor suppression, and disease control. However, existing methods typically rely on conventional network structures and fail to capture the polyadic interactions intrinsic to many real-world systems. To address this limitation, we study key node identification in hypergraphs, where higher-order interactions are naturally modeled as hyperedges. We propose a novel framework, \textbf{AHGA}, which integrates an \textbf{A}utoencoder for extracting higher-order structural features, a \textbf{H}yper\textbf{G}raph neural network-based pre-training module (HGNN), and an \textbf{A}ctive learning-based fine-tuning process. This fine-tuning step plays a vital role in mitigating the gap between synthetic and real-world data, thereby enhancing the model’s robustness and generalization across diverse hypergraph topologies. Extensive experiments on eight empirical hypergraphs show that AHGA outperforms classical centrality-based baselines by approximately 37.4\%. Furthermore, the nodes identified by AHGA exhibit both high influence and strong structural disruption capability, demonstrating its superiority in detecting multifunctional nodes.
\end{abstract}


\end{frontmatter}

%\linenumbers
\section{Introduction}
\makeatletter
\newcommand{\rmnum}[1]{\romannumeral #1}
\newcommand{\Rmnum}[1]{\expandafter\@slowromancap\romannumeral #1@}
\makeatother

Complex systems capture the intricate interaction patterns among entities in real-world scenarios, thereby advancing the understanding of dynamic mechanisms in systems such as social networks and power grids~\citep{barabasi2013network}. Key node identification aims to detect multifunctional nodes within a network using targeted strategies, thereby promoting effective information diffusion, strengthening network resilience, and suppressing the spread of rumors~\citep{albert2002statistical}.

Methods for identifying key nodes in conventional networks are generally categorized into two groups: heuristic approaches and data-driven algorithms. Heuristic methods evaluate node importance based on network topology, including degree centrality (DC), which captures local connectivity~\citep{bonacich1972factoring}, as well as betweenness centrality (BC) and closeness centrality (CC), both of which reflect global structural characteristics~\citep{freeman1977set,bavelas1950communication}. These approaches are widely adopted due to their simplicity, interpretability~\citep{freeman2002centrality}. To improve the precision of key node identification, data-driven techniques have been introduced. These include traditional machine learning models such as Least Squares Support Vector Machines and the Machine Learning Index framework~\citep{wen2018fast,yu2020identifying}, as well as deep learning models based on neural networks, including Graph Convolutional Networks (GCNs) and Graph Neural Networks (GNNs)~\citep{chen2015convolutional,kipf2016semi}. Owing to their powerful feature extraction capabilities, these advanced methods can effectively capture latent structural patterns associated with node characteristics and facilitate the training of downstream tasks~\citep{mohamadi2015trust}.

However, in real-world systems, entities often engage in complex higher-order interactions that go beyond simple dyadic relationships. Typical examples include information reinforcement in group-based social communications and synergistic effects in protein complexes~\citep{benson2016higher}. Hypergraphs, with their ability to represent multi-way relationships, offer a natural and powerful framework to model such higher-order interactions, thereby overcoming the structural limitations of conventional networks~\citep{battiston2020networks}. Against this backdrop,  identifying key nodes in hypergraphs has become increasingly important, prompting extensions of traditional heuristic algorithms, originally developed for conventional networks, into the hypergraph domain.  These include Degree Centrality (DC), Vector Centrality (VC), and Harmonic Closeness Centrality (HCC)~\citep{kepes2023critical, tudisco2021node}. In addition, various methods have exploited hypergraph-specific characteristics, such as the presence of hyperedges and the notion of higher-order distances, to develop tailored centrality metrics, including Hyperdegree Centrality (HDC), Higher-order Gravity Centrality (HGC), and Higher-order Distance Fuzzy Centrality (HDF)~\citep{zhang2025locating,xie2023vital}. Nevertheless, key node identification in hypergraphs continues to face notable challenges. \textbf{(i) Data-driven approaches remain underdeveloped in this domain}, indicating considerable potential for improving identification accuracy by leveraging hypergraph-specific structural features. \textbf{(ii) Both heuristic and data-driven methods often exhibit strong sensitivity to the underlying network characteristics}, resulting in limited robustness and generalizability of the identification outcomes.

To bridge the aforementioned gaps, we propose an innovative framework named AHGA, which comprises three key components, i.e., Autoencoder, Pre-Training, and Active Learning. The autoencoder leverages a Hypergraph Neural Network (HGNN) to capture higher-order topological information and generate high-quality node representations. The pre-training module further refines these initial features by combining HGNN with fully connected layers, ensuring strong baseline performance on synthetic datasets. The active learning module fine-tunes the basic model using representative nodes, thereby enhancing its performance in key node identification tasks on empirical hypergraphs. To validate the effectiveness of our proposed framework, we conduct extensive experiments on eight real-world hypergraph datasets, comparing AHGA with several benchmark methods. The results demonstrate that AHGA consistently outperforms existing approaches. Accordingly, the main contributions of this study are summarized as follows:

\begin{enumerate}[\textbullet]
    \item We employ hypergraph neural network techniques for key node identification, effectively overcoming the limitations of existing methods, including poor robustness and the inability to capture the higher-order structural information intrinsic to hypergraphs.
    
    \item We incorporate an active learning strategy to fine-tune the basic model, further improving its performance and scalability. As a result, AHGA exhibits strong practical applicability across a wide range of hypergraphs.

    \item By focusing on the identification of multifunctional nodes, AHGA surpasses traditional single-objective methods, as the selected nodes not only exhibit high influence but also play a critical role in maintaining network connectivity.
\end{enumerate}

The remainder of this paper is organized as follows: Section \ref{Related works} introduces the current research that is related to our work. The fundamental definitions of hypergraphs and the propagation model employed are introduced in Section \ref{Preliminary definition}. Section \ref{Methods} presents the proposed AHGA framework in detail. Section \ref{Baselines and Datasets} outlines the baseline methods used for comparison and the dataset employed in the experiment. Section \ref{Experiments} reports the evaluation metrics, experimental results, and corresponding analyses. Finally, Section \ref{Conclusions} concludes the study and discusses potential directions for future research.
\section{Related works} \label{Related works}

The problem of key node identification has traditionally been studied within the context of conventional networks. Existing solutions can be broadly categorized into heuristic methods and data-driven algorithms. Heuristic approaches, such as degree centrality~\citep{bonacich1972factoring}, betweenness centrality~\citep{freeman1977set}, and closeness centrality~\citep{bavelas1950communication}, evaluate node importance from different topological perspectives. Specifically, degree centrality reflects local influence by counting immediate connections, betweenness centrality captures a node’s control over global information flow through shortest paths, and closeness centrality measures a node’s overall efficiency in reaching all others in the network. Although these measures are intuitive and computationally tractable, they suffer from limitations such as sensitivity to neighbor quality, high computational cost, and reliance on network connectivity. In contrast, data-driven methods have demonstrated increasing potential in handling more complex and large-scale networks. Machine learning models like Least Squares Support Vector Machines (LS-SVMs) perform regression-based node ranking through quadratic optimization, while the Machine Learning Index (ML-Index) improves structural feature retrieval using hierarchical representations and error-bounded searches. Deep learning techniques further enhance these capabilities: Convolutional Neural Networks (CNNs) extract local structural features by encoding adjacency matrices, whereas Graph Convolutional Networks (GCNs) aggregate multi-hop neighbor information via spectral convolution operations. Recent variants such as infGCN~\citep{zhao2020infgcn}, which models influence propagation using attention mechanisms, and CGNN~\citep{zhang2022new}, which simulates dynamic spreading processes via coupled epidemic operators, significantly improve the extraction of nonlinear structural representations. However, despite these advancements, the application of data-driven methods to hypergraphs remains relatively unexplored. Effectively capturing node-level topological characteristics in the presence of complex higher-order interactions poses substantial challenges, underscoring the need for novel approaches tailored to hypergraph structures.

With the increasing complexity of networked systems, particularly with the emergence of hypergraphs that model higher-order interactions, there has been a growing interest in extending key node identification methods to hypergraph structures. Several heuristic approaches have been adapted to this context, including Degree Centrality (DC), Vector Centrality (VC), and Harmonic Closeness Centrality (HCC), which have been generalized to account for the unique characteristics of hypergraphs~\citep{kepes2023critical,tudisco2021node}. Beyond these adaptations, researchers have proposed novel centrality measures that explicitly leverage hypergraph-specific features such as hyperedges and higher-order distances. For instance, Hyperdegree Centrality (HEDC) evaluates a node’s significance based on its frequency of participation in hyperedges. Higher-order Gravity Centrality (HGC) combines node degree with higher-order distance metrics to provide a more nuanced assessment of influence. Higher-order Distance Fuzzy Centrality (HDC) further advances this line of work by incorporating fuzzy set theory and Shannon entropy to model the uncertainty and heterogeneity inherent in higher-order neighbor interactions~\citep{zhang2025locating,xie2023vital}. These metrics reflect an important shift toward capturing the rich structural semantics embedded in hypergraph-based systems.

\section{Preliminary definition}
\label{Preliminary definition}
\subsection{Definition of a hypergraph}
A hypergraph $G^H = (V, E)$ consists of a set of nodes $V = \{v_1, v_2, \ldots, v_N\}$ and a set of hyperedges $E = \{e_1, e_2, \ldots, e_M\}$, where $N$ and $M$ denote the numbers of nodes and hyperedges, respectively. Each hyperedge $e \in E$ is defined as a non-empty subset of $V$ with cardinality $|e| \geq 1$, representing higher-order interactions among the $|e|$ participating nodes.
In the context of a hypergraph, the adjacency matrix $A \in \mathbb{R}^{N \times N}$ encodes pairwise node relationships induced by shared hyperedges. Specifically, $A_{ij} = 1$ if nodes $v_i$ and $v_j$ both belong to at least one common hyperedge; otherwise, $A_{ij} = 0$. To characterize node-hyperedge affiliations, the incidence matrix $H \in \mathbb{R}^{N \times M}$ is introduced ~\citep{kermack1927contribution}, where $H_{ij} = 1$ if node $v_i$ is a member of hyperedge $e_j$, and $H_{ij} = 0$ otherwise. The node degree matrix is diagonal and represented by $K^V$, in which each diagonal element is given by $K^V_{ii}=\sum_{j=1}^{N}A_{ij}$. The node hyperdegree matrix $K^H$ is a diagonal matrix with entries $K_{ii}^H = \sum_{j=1}^{M} H_{ij}$, reflecting the connectivity of node $v_i$ across hyperedges. Similarly, the hyperedge size matrix $K^E$ is defined with diagonal entries $K_{jj}^E = \sum_{i=1}^{N} H_{ij}$, representing the number of nodes incident to hyperedge $e_j$.
\subsection{Definition of $s$-distance and $s$-line graph}
In a hypergraph, two hyperedges $e_p, e_q \in E$ are said to be $s$-adjacent if they share at least $s$ common nodes, i.e., $|e_p \cap e_q| \geq s$~\citep{aksoy2020hypernetwork}. An $s$-walk of length $l$ is defined as a sequence of hyperedges $\mathcal{W}_i^l = \{e_{i0}, e_{i1}, \ldots, e_{il}\}$ such that each pair of consecutive hyperedges $(e_{ik}, e_{i(k+1)})$ satisfies the $s$-adjacency condition. If the sequence contains no repeated nodes, it is termed an $s$-path. The $s$-distance between two hyperedges $e_p$ and $e_q$, denoted by $d_s^e(p, q)$, is defined as the length of the shortest $s$-path connecting them. If no such path exists, we set $d_s^e(p, q) = \infty$.
Given two nodes $v_i$ and $v_j$ belonging to hyperedges $e_p$ and $e_q$, respectively, the $s$-distance $d_s^v(i,j)$ between them is given as follows:
\begin{equation}
    d_s^v(i,j) =\begin{cases} 
1 & \text{if } p = q, \\
\displaystyle\min_{e_p \in E_i, e_q \in E_j} d_s^e(p,q) + 1 & \text{otherwise},
\end{cases}
\end{equation}
where $E_i$ denotes the set of hyperedges that contain node $v_i$.
Therefore, we define the $s$-distance matrix of the hypergraph as:
\begin{equation}
D_S = \{ d_s^v(i, j) \}_{N \times N}.
\end{equation}

The $s$-diameter of the hypergraph, denoted by $d_s^G$, is then calculated as:
\begin{equation}
d_s^G = \max_{i,j} d_s^v(i, j).
\end{equation}

The $s$-line graph of a hypergraph, denoted as $L(G^H_s) = (V^L_s, E^L_s)$, is constructed by representing each hyperedge $e \in E$ of the original hypergraph $G^H = (V, E)$ as a node $v_e \in V^L_s$ in $L(G^H_s)$. An edge exists between two nodes $v_{e_p}, v_{e_q} \in V^L_s$ if and only if the corresponding hyperedges $e_p, e_q \in E$ are $s$-adjacent, that is, $|e_p \cap e_q| \geq s$.
\subsection{Hypergraph-Based SIR Model}
In the hypergraph-based Susceptible-Infected-Recovered (SIR) model, each node can exist in one of three states: Susceptible (S), Infected (I), or Recovered (R)~\citep{suo2018information}. An infected node (I) can transmit the infection to its susceptible neighbors with probability $\beta$. Concurrently, each infected node has a probability $\gamma$ of transitioning to the recovered state (R), after which it remains permanently immune. To reflect realistic scenarios where an individual may interact with only one group within a short time interval, we assume that, at each discrete time step, an infected node can activate a single hyperedge~\citep{xie2023vital}. The detailed spreading dynamics under this constraint are described as follows:
\begin{itemize}[]
    \item Initially, a seed node is set to the I state, while all other nodes are initialized in the S state.
    
    \item At each discrete time step $t$, all infected nodes are identified. For each infected node $v_i$, we use $E_i$ to represent the hyperedge set that contains node $v_i$. Then $e_j \in E_i$ is randomly selected, and every susceptible node within $e_j$ is independently infected by $v_i$ with probability $\beta$. Simultaneously, node $v_i$ recovers and transitions to the R state with probability $\gamma$.

    \item The spreading process continues iteratively until no infected nodes remain in the hypergraph.
\end{itemize}
\begin{figure}[] 
    \centering
    \includegraphics[width=1\linewidth]{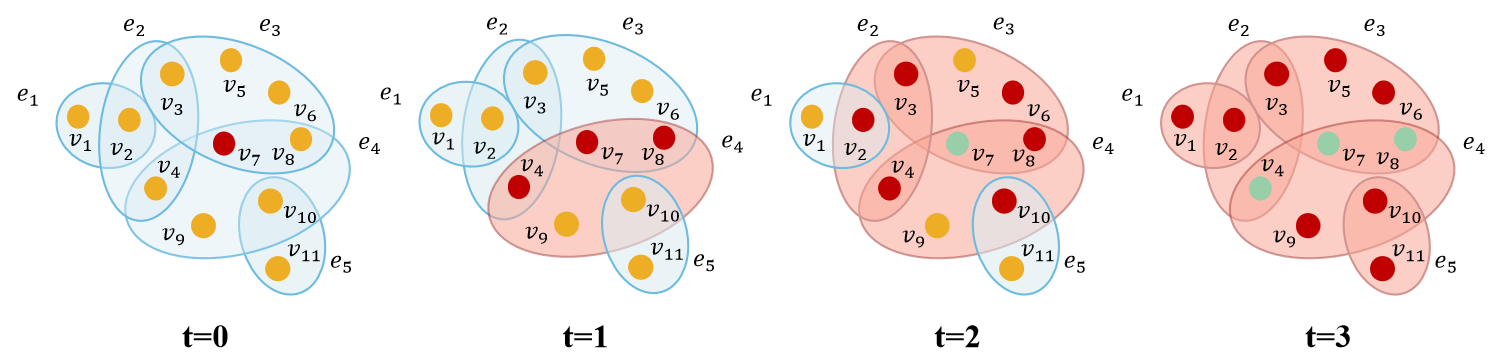}
    \caption{Illustration of the hypergraph-based SIR spreading process. Node colors indicate their dynamic states, i.e., orange for susceptible (S), red for infected (I), and green for recovered (R). }
    \label{figure1}
\end{figure}
For each hypergraph used in the numerical experiments, the infection probability is set to a value $\beta_0 \gtrsim \beta_c$, where $\beta_c$ denotes the critical threshold of the spreading process. The specific values of $\beta_0$ for all hypergraphs are listed in Table~\ref{table1}. In addition, Figure~\ref{figure1} provides a visual example of the SIR spreading process on a hypergraph. Specifically, node $v_7$ is initialized as the infected seed at time step $t = 0$. The set of hyperedges containing $v_7$ is $E_7 = \{e_3, e_4\}$. At time step $t = 1$, hyperedge $e_4$ is randomly selected, resulting in the infection of nodes $v_4$ and $v_8$. At time step $t = 2$, the contagion propagates through hyperedges $e_2$ and $e_3$, infecting nodes $v_2$, $v_3$, $v_6$, and $v_{10}$, while node $v_7$ transitions to the R state. By time step $t = 3$, the infection has spread across the entire hypergraph, i.e., all previously susceptible nodes become infected, and nodes $v_4$ and $v_8$ recover. 

To obtain the ground-truth influence of each node, we individually set every node as the initial seed and simulate the hypergraph-based SIR model. The final number of infected individuals, including those who recover, is recorded as the influence value of the corresponding node. A node ranking method is considered to exhibit high performance in key node identification if it effectively prioritizes nodes with high spreading influence.

\section{Methods}\label{Methods}
\begin{figure}[] 
    \centering
    \includegraphics[width=1\linewidth]{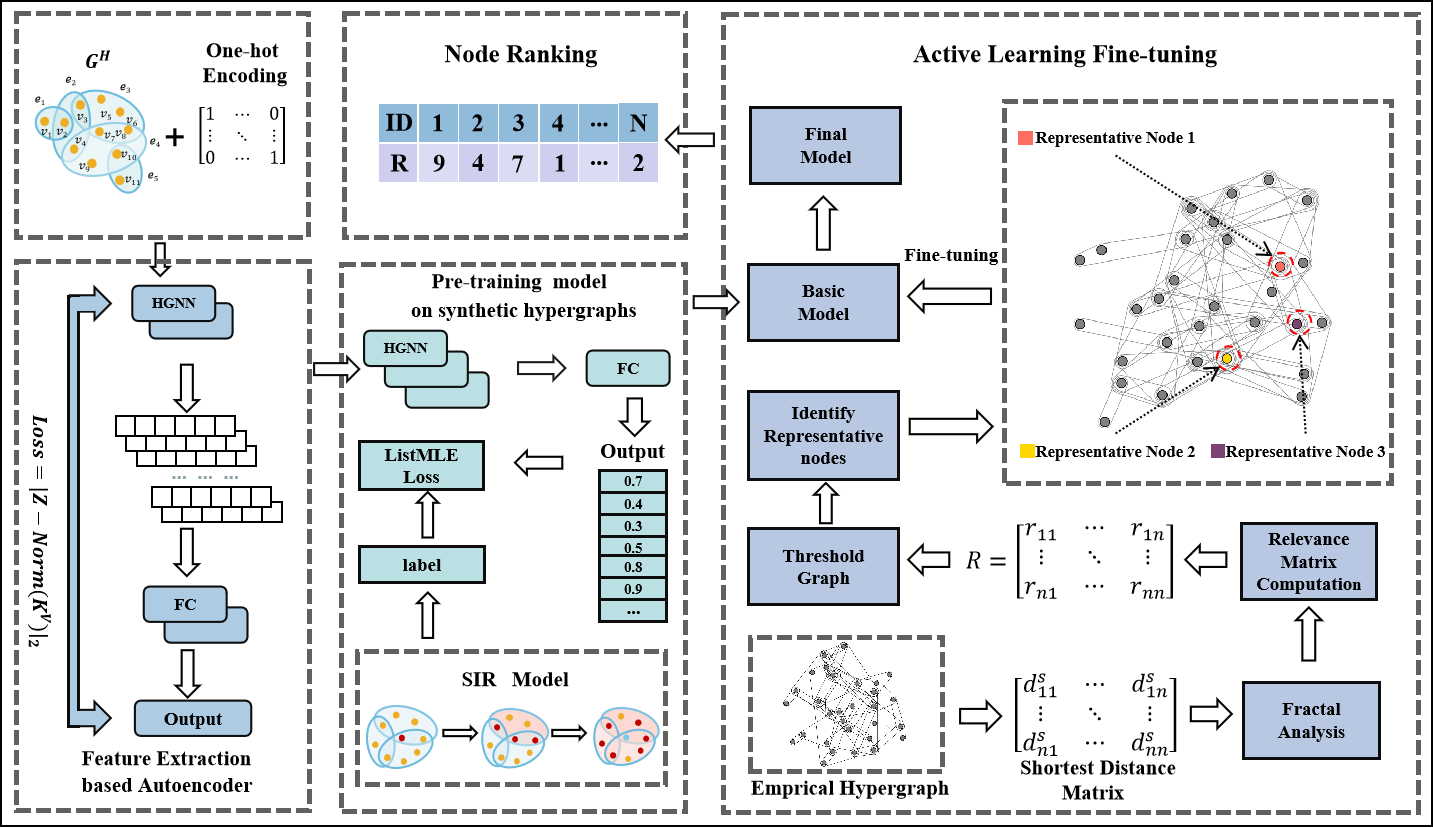}
    \caption{Framework of the AHGA. The hypergraph structure and one-hot encoded node features are fed into an autoencoder, which leverages a hypergraph neural network (HGNN) to learn latent node representations. These embeddings serve as input to the task training module, producing a basic model that outputs node importance scores. The fine-tuning component further refines this model via active learning, resulting in the final optimized model.}
    \label{figure2}
\end{figure}

We propose a three-stage learning framework, AHGA, for identifying important nodes in a hypergraph. The framework comprises: (1) an \textbf{A}utoencoder, (2) a pre-training model based on \textbf{H}yper\textbf{G}raph neural network (HGNN), and (3) an \textbf{A}ctive learning. AHGA is designed to effectively capture higher-order relational structures inherent in hypergraph data while improving the model’s generalization capability. An overview of the proposed framework is illustrated in Figure~\ref{figure2}.
\subsection{Autoencoder-based feature extraction}
To capture the higher-order relationships between nodes and hyperedges in hypergraphs, we design an autoencoder architecture that integrates an HGNN into the encoding process. The autoencoder encodes topological features into compact, low-dimensional latent representations while preserving essential structural information, and subsequently reconstructs them to approximate the original input. By minimizing the loss of reconstruction, the model learns to retain the most informative patterns embedded in the hypergraph structure. Incorporation of HGNN into the encoder further enhances the ability of the model to extract complex higher-order dependencies inherent in the hypergraph 
data~\citep{hinton2006reducing}.

The encoder is composed of three layers of HGNNs, which take as input the hyperedge size matrix $K^E$, the incidence matrix $H$, the one-hot encoded node features, and the node hyperdegree matrix $K^H$. Each HGNN layer operates in two stages. In the first stage, node features are projected onto the hyperedges by aggregating information from all incident nodes, as formulated by the following equation:
 \begin{equation}
Z_{e}^{(l)}=\sigma\left((K^{E})^{-1}H^{T}WX^{(l)}\Theta_{e}^{(l)}\right),
 \end{equation}
where $W$ is a randomly initialized matrix whose elements are independently sampled from a uniform distribution over the interval $[0,1]$, $Z_e^{(l)}$ denotes the hyperedge features at the $l$-th layer, $X^{(l)}$ is the node feature matrix, $\Theta$ represents the learnable parameters, and $\sigma(\cdot)$ denotes the activation function, which is set to ReLU in our implementation. Next, the hyperedge features, i.e., now enriched by the aggregated information from nodes, are integrated back into the corresponding nodes to update their representations.
 \begin{equation}
     X^{(l + 1)}=\sigma\left((K^{H})^{-1/2}HZ_{e}^{(l)}\Theta_{v}^{(l)}\right).
\end{equation}

Therefore, the complete mathematical formulation for this process is as follows:
\begin{equation}\small
X^{(l + 1)} = \sigma \left( (K^H)^{-1/2} H W  (K^E)^{-1} H^T \\ (K^H)^{-1/2} X^{(l)} \Theta^{(l)} \right).
\end{equation}

In the encoder, we denote the initial node feature matrix as $X^0 \in \mathbb{R}^{N \times N}$. The learnable parameter matrix at the first layer is $\Theta^0 \in \mathbb{R}^{N \times d}$, where $d$ is an integer divisible by 4. After the first HGNN layer, the node representation is updated to $X^1 \in \mathbb{R}^{N \times d}$. The second-layer parameter matrix is $\Theta^1 \in \mathbb{R}^{d \times \frac{d}{4}}$, and the final output of the encoder is denoted by $X^2 \in \mathbb{R}^{N \times \frac{d}{4}}$.

In the decoder, we utilize a multi-layer perceptron (MLP) consisting of two fully connected (FC) layers to reconstruct the high-dimensional features from the low-dimensional representations learned by the encoder. The transformation at each layer is defined as follows: 
\begin{equation}
Y^{i + 1}=Y^iW^i_{MLP}+b^i,
\end{equation}
where $Y^i$ denotes the node representation at the $i$-th FC layer, $W^i_{\text{MLP}}$ represents the trainable weight matrix, and $b^i$ is the corresponding bias term. The input to the decoder is denoted by $Y^0 \in \mathbb{R}^{N \times d}$, which corresponds to the encoder’s output. The final decoder output is given by $Y^2 \in \mathbb{R}^{N \times 1}$. The loss function used to optimize the node feature reconstruction is defined as follows:
\begin{equation}
L_1=\left\lVert Z - Norm(K^V)\right\rVert_2,
\end{equation}
where $\left\lVert \cdot \right\rVert_2$ denotes the Euclidean norm, $Z$ is the reconstructed output of the model, $Norm$ represents a normalization operation, and $K^V$ is the diagonal matrix of node degrees. We select node degree as the reconstruction target because it efficiently captures local connectivity patterns and node influence within the hypergraph, while offering low computational complexity. This makes it particularly suitable for feature extraction in autoencoder-based models. By minimizing this loss, the model is guided to learn feature representations that are aligned with node connectivity patterns, which are essential for tasks such as key node identification.

\subsection{Pre-training model}
Key node identification is formulated as a regression task. The node features extracted from the autoencoder are fed into a pre-training model composed of a three-layer HGNN followed by a single FC layer. The stacked HGNN layers enable the model to capture higher-order structure across both local and global contexts. The final FC layer integrates features from all dimensions and outputs a scalar score representing the importance of each node.

The node influence labels for each hypergraph are derived from simulations based on the hypergraph-structured SIR model. For model training, we adopt the Adam optimizer with a learning rate of 0.01 and use ListMLE as the loss function. Unlike mean squared error (MSE), which evaluates individual predictions, ListMLE is specifically designed for learning-to-rank tasks by optimizing the entire ranking list. Its core principle is to maximize the likelihood of the ground-truth ranking order, thereby enabling more effective training of the ranking model and achieving improved ranking performance. The corresponding mathematical formulation is given by:
\begin{equation}
L_2= -\log P(\pi_y \mid y'),
\end{equation}
where $y'$ denotes the predicted scores output by the FC layer, $\pi_y$ represents the ground-truth ranking derived from the true labels $y$, and $P$ denotes the probability distribution defined by the Plackett–Luce model~\citep{plackett1975analysis}.

\subsection{Active learning fine-tuning}

As demonstrated in prior studies, model performance can vary significantly across different hypergraph structures~\citep{kipf2016semi}. To improve generalization and enhance overall performance, we incorporate active learning strategies to select representative nodes from each empirical hypergraph, which are then used to fine-tune the base model. To obtain representative nodes, we first construct a node similarity matrix based on both global and local fractal dimensions, which captures the structural similarity between nodes. Subsequently, subgraphs are extracted from this similarity matrix using a tailored strategy, and the selected representative nodes within these subgraphs are then utilized for fine-tuning.

Specifically, we first compute the $s$-distance matrix $D_S$ and determine the $s$-diameter $d_s^G$ of the hypergraph. Based on this, we estimate both the global fractal dimension $d_f$ and the local fractal dimension $d_l$. The global fractal dimension $d_f$ characterizes the structural complexity of the hypergraph~\citep{clauset2009power}, which is evaluated by quantifying the number of boxes $B$ required to cover all nodes in the hypergraph, with each box containing nodes within a distance less than or equal to a given size $r_B$  $(r_B \in [1, d_s^G])$.
In fractal theory, the minimum number of boxes $B$ required to cover the entire hypergraph and the box size $r_B$ follow a power-law relationship. By taking the natural logarithm of both variables, this relationship can be linearized as follows:
\begin{equation}
    \ln B(r_B)=C_1-d_f\cdot\ln r_B,
\end{equation}
where $C_1$ is a constant and $B(r_B)$ indicates the number of boxes that cover the entire hypergraph, determined by $r_B$. Rearranging the equation allows for explicit computation of $d_f$, i.e.,
\begin{equation}
    d_f = \frac{\ln B(r_B)-C_1}{\ln(r_B)}.
\end{equation}

Since varying the box size $r_B$ results in different corresponding values of $B$, the global fractal dimension $d_f$ can be estimated by performing a least squares method on the log-transformed relationship between $B$ and $r_B$~\citep{jin2014scalable}. 

Inspired by the global fractal dimension derived from the box-covering method, we propose a circle-covering approach to measure the local fractal dimension, focusing on individual nodes. Centered at a given node $v_i$, we define a circle with radius $r_l$ ($r_l \in [1, \frac{d_s^G}{2}]$), which includes any node $v_j$ satisfying $d^s_v(i,j) \leq r_l$. The set of nodes within this circle is denoted as $N_i^{r_l}$. Unlike the global fractal dimension, the cardinality of $N_i^{r_l}$ increases with $r_l$, allowing us to express a direct linear relationship between these two variables:
\begin{equation}
    d_l^{(v_i)}=\frac{|N_i^{r_l}|-C_2}{r_l}.
\end{equation}

Then, the least squares method can be directly applied to fit $d_l^{(v_i)}$, which quantifies the structural complexity in the local neighborhood centered at node $v_i$.

Based on $d_l^{(v_i)}$, we compute the proportion $P_{i}^{K^V}$ of nodes within the set $V_k$, which comprises all nodes with degree $K^V$ that appear in the $r_l$-hop neighborhood $N^{r_l}_i$ of the central node $v_i$.  $P_{i}^{K^V}$ integrates the structural information of $V_k$ to the center node $v_i$:
\begin{equation}
P_{i}^{K^V}=\frac{\sum_{v_0\in V_{k}} d_l^{(v_{0})}}{\sum_{v_j\in\mathcal N^{r_l}_i}d_l^{(v_j)}}.
\end{equation}

Subsequently, we generate the node similarity matrix $S$ based on $d_f$ and $P_i^{K^V} $, $i \in[1,N]$,  where each element $S_{ij}$ in $S$ is represented as follows:

\begin{equation}
    S_{ij}=\sum_{K^V=0}^{max(K^V_i,K^V_j)}\frac{\left(\frac{P_{i}^{K^V}}{P_{j}^{K^V}}\right)^{d_f}-\frac{P_{i}^{K^V}}{P_{j}^{K^V}}}{1-d_f}({P_{j}^{K^V}\neq0)},
\end{equation}
where $S_{ij}$ denotes the similarity from node $v_i$ to $v_j$. Thus, it is necessary to further capture the mutual relationship between nodes $v_i$ and $v_j$, each element $r_{ij}$ in the final relevance matrix $R$ is defined as below:
% $d_f$, as the power-law exponent, directly influences the quantification of node relationships. 
\begin{equation}
    r_{ij} = S_{ij}+S_{ji},
\end{equation}
where a larger value of $r_{ij}$ indicates that $v_i$ and $v_j$ have a higher structural similarity.

Finally, we construct a subgraph $G_a$~\citep{leskovec2007cost} to identify a set of representative nodes. Specifically, we begin with an empty graph and iteratively add each node pair $(v_i, v_j)$ to $G_a$ if their similarity score $r_{ij}$ exceeds a predefined threshold $\theta$. In this constructed subgraph, a higher node degree reflects a greater number of similar nodes, as edges are formed based on the similarity matrix. To extract representative nodes, we adopt an iterative pruning strategy: in each iteration, the node with the highest degree is selected as a representative, and both the selected node and its neighbors are removed from $G_a$. The degrees of the remaining nodes are then updated accordingly. This process repeats until the desired number of representative nodes, denoted by $N_{\text{rep}}$ (set to 10), is reached. The selected representative nodes will compose the new training set to fine-tune the pre-trained model. By leveraging active learning techniques, the model adapts to node identification tasks across diverse hypergraphs in a cost-effective manner.
\section{Baselines and Datasets}
\label{Baselines and Datasets}
\subsection{Baselines}
To evaluate the effectiveness of the proposed AHGA framework, we compare it against a set of representative baseline methods, including Degree Centrality (DC), Hyperedge Degree Centrality (HEDC), Vector Centrality (VC), Harmonic Closeness Centrality (HCC), and $s$-distance based fuzzy centrality (HDF). We provide detailed descriptions of these baseline methods as follows.

\textbf{Degree Centrality (DC)~\citep{zhang2017degree}:} Degree centrality is one of the most fundamental measures in network analysis, quantifying the importance of a node based on the number of direct connections it has. The degree centrality of node $v_i$ is computed as:
\begin{equation}
K_i^V = \sum_{j=1}^{N} A_{ij},
\end{equation}
where $A_{ij}$ denotes the $(i,j)$-th entry of the adjacency matrix $A$. A higher degree value indicates that the node is more important within the hypergraph.

\textbf{Hyperedge Degree Centrality (HEDC)~\citep{hu2021aging,wang2010evolving}:} HEDC evaluates node importance by incorporating the degree of hyperedges to which nodes belong. Specifically, the calculation is based on the $1$-line graph of the hypergraph, denoted as $L(G^H_1) = (V^L_1, E^L_1)$. The degree of a hyperedge $e_m$ in the line graph is defined as:

\begin{equation}
    C_{HEDC}^e(m) = \sum_{g = 1}^{M} A_{mg}^L,
\end{equation}
where $A^L$ is the adjacency matrix of the line graph $L(G^H_1)$, with $A_{mg}^L = 1$ if and only if hyperedges $e_m$ and $e_g$ share at least one common node, i.e., are adjacent, and $0$ otherwise.

Then, the HEDC value of a node $v_i$ is computed by aggregating the degree contributions of its incident hyperedges as follows:

\begin{equation}
    C_{HEDC}^v(i) = \sum_{m = 1}^{M} H_{im} \frac{C_{HEDC}^e(m)}{K^E_m},
\end{equation}
where $K^E_m$ is the size of hyperedge $e_m$. A higher HEDC value indicates that the node is considered more important in the hypergraph.

\textbf{Vector Centrality (VC)~\citep{kovalenko2022vector}:} VC assesses node importance by leveraging eigenvectors, where the centrality scores are first computed for hyperedges. Each hyperedge score is then evenly distributed among its associated nodes to obtain the final node-level centrality. The hypergraph is first transformed into its corresponding line graph $L(G^H_1) = (V^L_1, E^L_1)$, where eigenvector-based scores are computed for all hyperedges. Let $C^e_{VC}(m)$ denote the VC score of hyperedge $e_m$, where $m = 1, \ldots, M$. For a node $v_i \in e_m$, we define $c^v_{iK^E_m}$ as the portion of the VC score contributed by hyperedge $e_m$ with size $K^E_m$ to node $v_i$, where $K^E_m \in [2, \ldots, K^E_{\max}]$ and $K^E_{\max} = \max_m K^E_m$. Therefore, the node-level VC score could be represented as 
\begin{equation}
    \vec{c}_i = (c_{i2}, \ldots, c_{iK^E_{max}})^T \in \mathbb{R}^{K^E_{max}-1},
\end{equation}
where each element is computed as $c_{iK^E_m} = \frac{1}{K^E_m} \sum_{i \in e_m} C^e_{VC}(m)$, meaning the score of a hyperedge is evenly distributed among its constituent nodes. The final VC score of node $v_i$ is obtained by summing over all components of $\vec{c}_i$: 

\begin{equation}
    C^v_{VC}(i) = \sum^{K^E_{max}}_{K^E_m = 2} c_{iK^E_m}.
\end{equation}

Therefore, a higher value of $C^v_{VC}(i)$ indicates that node $v_i$ plays a more important role within the hypergraph.

\textbf{Harmonic Closeness Centrality (HCC)~\citep{aksoy2020hypernetwork}:}  HCC measures the importance of a hyperedge based on its average $s$-distance to all other hyperedges, with the intuition that a hyperedge closer to others plays a more central role in the hypergraph. The computed HCC score of each hyperedge is evenly assigned to its incident nodes, thereby reflecting node importance from a global structural perspective. HCC is calculated on the 1-line graph $L(G^H_1) = (V^L_1, E^L_1)$, and its formulation is given by:

\begin{equation}
C_{\text{HCC}}^e(g) = \frac{1}{M - 1} \sum_{\substack{e_g,e_q \in E_s \\ g \neq q}} \frac{1}{d_s^e(g, q)},
\end{equation}
where $d_s^e(g, q)$ represents the $s$-distance between hyperedges $e_g$ and $e_q$. If $d_s^e(g, q) = \infty$, then $\frac{1}{d_s^e(g, q)} = 0$.
Therefore, the HCC score of a node $v_i$ is given as follows:

\begin{equation}
    C_{HCC}^{v}(i)=\sum_{m = 1}^{M} H_{im} \frac{C_{HCC}^{e}(m)}{K^E_m}.
\end{equation}

In the experimental setup, we set $s = 1$ to compute the HCC scores of nodes. Nodes with higher HCC values are considered more important, as they exhibit greater proximity to other nodes in the hypergraph.

\textbf{Higher-order Distance based fuzzy centrality (HDF)~\citep{zhang2025locating}:} HDF quantifies node importance by considering the collective influence of its higher-order neighbors through the fuzzy set theory and Shannon entropy. The method assumes that nodes closer to a central node $v_i$ exert a stronger influence on it. To model this, a hypersphere centered at $v_i$ is constructed, with its radius determined by the $s$-distance. The influence of each surrounding node is then estimated based on its distance to $v_i$.
For a node at an $s$-distance $l_i^s$ from the central node $v_i$, its influence is defined via a fuzzy membership function:

\begin{equation}
    x\left(l_{i}^{s}\right)=\exp \left(-\frac{\left(l_{i}^{s}\right)^{2}}{\left(L_{i}^{s}\right)^{2}}\right),
\end{equation}
where $L_{i}^{s} = \left\lceil \frac{z_{i}^{s}}{r} \right\rceil$, with $z_{i}^{s}$ denoting the maximum $s$-distance from node $v_i$ to all other nodes, and $r$ being an adjustable parameter. A smaller value of $r$ results in a larger effective radius, thereby including more neighboring nodes in the influence region of $v_i$.

Assume that the number of nodes at a distance of $l_{i}^{s}$ from node $v_i$ is $n(l_{i}^{s})$, then the number of fuzzy nodes at a distance of $l_{i}^{s}$ is $f\left(l_{i}^{s}\right)=n\left(l_{i}^{s}\right) x\left(l_{i}^{s}\right) x\left(l_{i}^{s}\right)$, and the total number of fuzzy nodes within the sphere is: $F(L_{i}^{s})=\sum_{l_{i}^{s}=1}^{L_{i}^{s}} f(l_{i}^{s})$. Based on this, we can calculate the proportion of nodes with the shortest s-distance of $l_i^s$ from node $v_i$ as $p\left(l_{i}^{s}\right)=\frac{1}{e} \frac{f\left(l_{i}^{s}\right)}{F\left(L_{i}^{s}\right)}$. Here, $\frac{1}{e}$ is used as a scaling factor to refine the probability to the range of $[0,\frac{1}{e}]$ so as to characterize the node influence using Shannon entropy. Thus, we can obtain the local fuzzy centrality of node $v_i$ as:
\begin{equation}
C_{HDF}(i)=\frac{\sum_{s = 1}^{s_{m}} C_{HDF}^{s}(i)}{s_{m}},
\end{equation}
where $C_{\text{HDF}}^s(i) = \sum_{l_i^s = 1}^{L_i^s} \frac{-p(l_i^s) \ln(p(l_i^s))}{(l_i^s)^2}$ represents the HDF score at distance $s$, and $s_m$ ($1 \leq s_m \leq s_M$) is a tunable parameter. This formulation integrates the entropy-weighted influence of multi-scale neighborhoods to comprehensively measure the importance of a node.

\subsection{Datasets}

We utilize synthetic hypergraphs as the training and validation sets, and employ eight empirical hypergraphs as the test set to evaluate the model’s performance. The details of the synthetic and empirical hypergraphs are provided as follows.

\subsubsection{Synthetic hypergraphs}

We employ three types of synthetic uniform hypergraphs as training and validation sets, i.e., Erdős-Rényi Hypergraph (ERH)~\citep{surana2022hypergraph}, Watts-Strogatz Hypergraph (WSH)~\citep{watts1998collective}, and Scale-Free Hypergraph (SFH)~\citep{feng2024hyper}. In all cases, uniform indicates that all hyperedges have the same cardinality. Each hypergraph consists of 1,000 nodes and 1,000 hyperedges. For ERH and WSH, the average hyperedge degree is set to 3, with a fixed hyperedge size of 3. For SFH, the average hyperedge degree is 10, and the hyperedge size is fixed at 5. The construction procedures for each hypergraph type are detailed below.

\textbf{ERH:} The ERH model constructs a uniform hypergraph $G^H = (V, E)$ through a random generation mechanism. Starting with an empty hyperedge set $E$, each iteration generates a candidate hyperedge $e$ by randomly selecting $K^E$ distinct nodes. If $e$ is not already present in $E$, it is added; otherwise, the selection is repeated. This process continues until $M$ unique hyperedges are obtained.

\textbf{WSH:} The WSH model generates a uniform hypergraph $G^H = (V, E)$ with small-world properties by combining local regularity with random perturbation. The process begins by constructing a ring-based regular hypergraph, where each node participates in hyperedges formed with its local neighbors within a fixed radius determined by $K^E$. For example, if $K^E = 3$, the initial hyperedges are sequentially formed as $e_1 = \{v_1, v_2, v_3\}$, $e_2 = \{v_2, v_3, v_4\}$, and so on, until $M$ hyperedges are generated. Subsequently, each hyperedge undergoes a rewiring step: with probability $p = 0.5$, each node in $e$ is replaced by a randomly selected node from $V \setminus e$. If the resulting hyperedge does not already exist in $E$, it is accepted; otherwise, the replacement is repeated. This process ensures that all hyperedges are perturbed once, embedding randomness into the regular structure.

 \textbf{SFH:} The SFH model generates a uniform hypergraph $G^H = (V, E)$ with a heterogeneous degree distribution. First, a sequence of node degrees $\{K^V_1, K^V_2, \ldots, K^V_N\}$ is drawn from a power-law distribution defined by $p(K^V) \propto (K^V)^{-\gamma}$, where $\gamma = 2$. The hyperedge set $E$ is initially empty. Each node $v_i$ is assigned a selection probability $p_i$ generated by $p(K^V)$. To construct a hyperedge, we initialize an empty set $e$ and iteratively add nodes to it based on their probabilities $p_i$ until $K^E$ unique nodes are included. If the constructed hyperedge does not already exist in $E$, it is added. This process repeats until the total number of hyperedges reaches $M$.

\subsubsection{Empirical hypergraphs}
\begin{itemize}
    \item \textbf{Senate-Comm \citep{chodrow2021generative}:} Nodes correspond to members of U.S. Senate committees, and each hyperedge links all members serving on the same committee.
    \item \textbf{Algebra \citep{amburg2020fair}:} Nodes represent users on \url{mathoverflow.net}, and each hyperedge connects all users who provided answers or comments on the same algebra-related question.
    \item \textbf{Rest-Rev \citep{amburg2020fair}:} Nodes represent Yelp users, and each hyperedge connects users who reviewed the same type of restaurants within a one-month period.
    \item \textbf{Geometry \citep{amburg2020fair}:} Nodes correspond to users on \url{mathoverflow.net}, with each hyperedge linking users who answered or commented on the same geometry-related question.
    \item \textbf{Music-Rev \citep{ni2019justifying}: }Nodes are Amazon users, and hyperedges connect the set of users who review the same category of music within one month.
    \item \textbf{House-Com \citep{chodrow2021generative}: }Nodes represent members of the U.S. House of Representatives committees, and hyperedges connect members belonging to the same committee.
    \item \textbf{Email-Enron \citep{Benson-2018-simplicial}: }Nodes are the email addresses of Enron employees, and hyperedges connect the senders and receivers of the same email.
    \item \textbf{Email-W3C \citep{amburg2021planted}: }Nodes are the email addresses on the W3C mailing list, and hyperedges connect the senders and receivers of the same email.
\end{itemize}
\begin{table}[]
\centering
\caption{Topological characteristics of the Datasets, in which we show the number of nodes ($N$), number of hyperedges (M), average node degree ($\langle K^V \rangle$), average node hyperdegree ($\langle K^H \rangle$), average hyperedge size ($\langle K^E \rangle$), coefficient of variation of node degree distribution ($CV(K^V)$) and the infection probability ($\beta_0$).}
\vspace{6pt}
\renewcommand{\arraystretch}{1}
\setlength{\tabcolsep}{6pt}
\begin{tabular}{cccccccc}
    \toprule
    Hypergraphs & $N$ & $M$ & $\langle K^V \rangle$ & $\langle K^H \rangle$ & $\langle K^E \rangle$ & $CV(K^V)$ & $\beta_0$ \\
    \midrule
    % \multirow{3}{*}{Synthetic} 
    WSH & 1000 & 1000 & 5.474 & 3.06 & 3.00 &0.48 &0.020  \\
    ERH & 1000 & 1000 & 5.982 & 3.141 & 3.00 & 0.58&0.030\\
    SFH & 1000 & 1000 & 38.874 & 10.905 & 5.00 &1.88 &0.008 \\
    %\midrule
    % \multirow{8}{*}{Empirical} 
    Senate-Com & 282 & 315 & 100.82 & 18.85 & 17.53 &0.44 &0.036 \\
    Algebra & 423 & 1268 & 78.90 & 19.53 & 6.52 & 0.87 &0.198\\
    Rest-Rev & 505 & 601 & 8.14 & 8.14 & 7.66 & 0.75 &0.026\\
    Geometry & 580 & 1193 & 164.79 & 21.52 & 10.47 &0.74 & 0.040\\
    Music-Rev & 1106 & 694 & 167.88 & 9.49 & 15.13 & 0.64&0.012 \\
    House-Com & 1290 & 341 & 195.58 & 9.17 & 35.25 &0.54 &0.012 \\
    Email-Enron & 2807 & 5000 & 20.41 & 4.62 & 4.985 &1.48 &0.014 \\
    Email-W3C & 5601 & 6000 & 3.53 & 2.39 & 2.223 & 3.66& 0.300\\
    \bottomrule

    \label{table1}
\end{tabular}
\end{table}

\begin{figure}[] 
    \centering
    \includegraphics[width=1\linewidth]{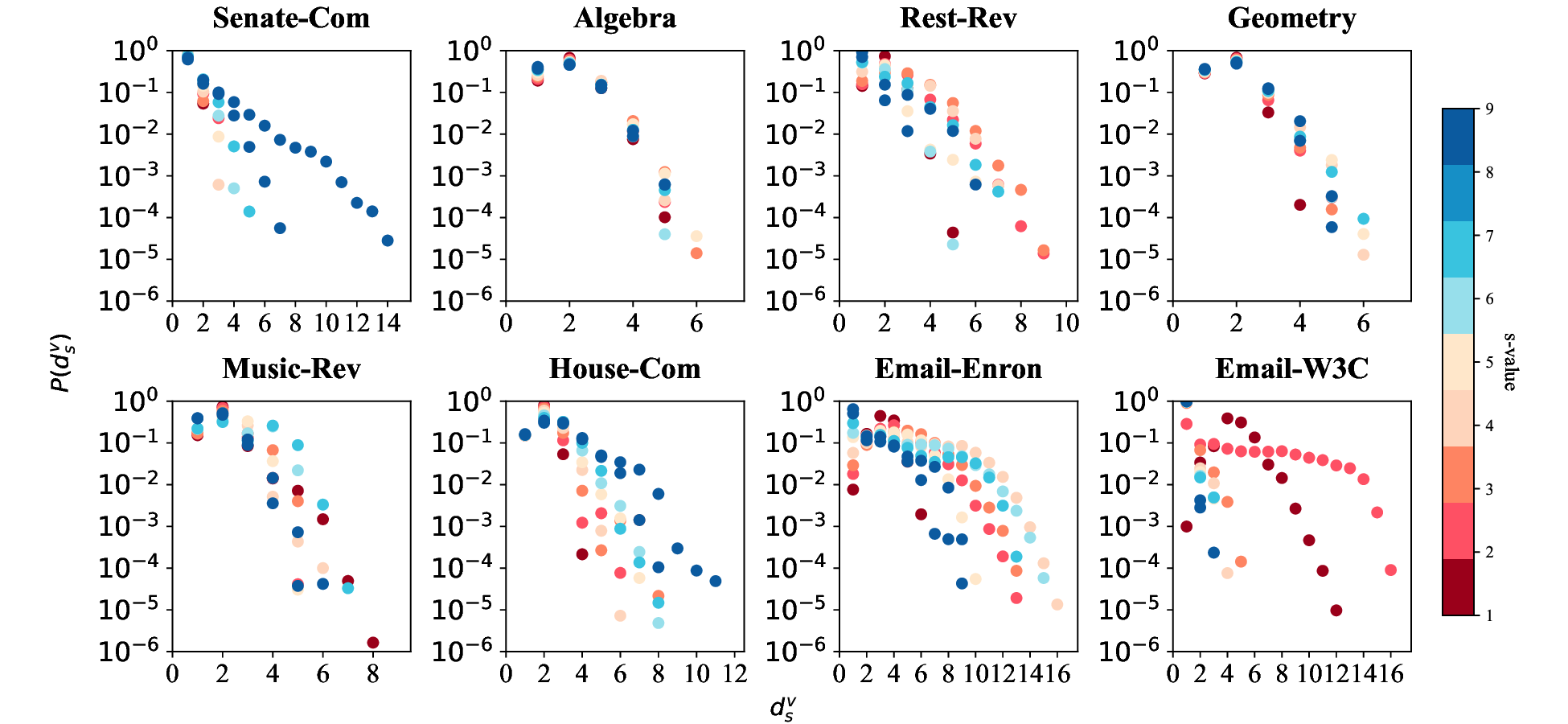}
    \caption{Distribution of $s$-distances ($d^v_s$) between nodes in the hypergraphs for different empirical hypergraphs, with $s$ ranging from 1 to 9. Scatter point colors range from light green to blue, representing the increasing value of $s$, as indicated by the color bar.}
    \label{fig4}
\end{figure}

Table~\ref{table1} summarizes key statistics for both synthetic and empirical hypergraphs, including the number of nodes ($N$), number of hyperedges ($M$), average node degree ($\langle K^V \rangle$), average node hyperdegree ($\langle K^H \rangle$), average hyperedge size ($\langle K^E \rangle$), coefficient of variation of node degree distribution ($CV(K^V)$) and the baseline infection probability ($\beta_0$). Furthermore, Figure~\ref{figure4} presents the distribution of $s$-distances $d^v_s$ across different datasets, with $s$ ranging from 1 to 9. For hypergraphs such as Email-Enron and Email-W3C, we observe that the $s$-distances corresponding to smaller $s$ values (i.e., $s \leq 5$) tend to be larger, as indicated by the warmer colors (red to yellow). In contrast, the $s$-distances for larger $s$ values (i.e., $s > 5$) are relatively shorter, as shown by the cooler blue tones. Conversely, in hypergraphs such as Senate-Com and House-Com, the pattern is reversed, where smaller $s$ values correspond to shorter distances and larger $s$ values to longer ones. This observation provides empirical guidance for selecting the appropriate $s$ value in the active learning component of our framework.

\section{Experiments}\label{Experiments}
To validate the effectiveness of our proposed method, we design three evaluation metrics. Specifically, Kendall’s $\tau$ coefficient and Rank Overlap are used to assess the capability of identifying influential nodes, while $s$-efficiency is employed to measure the method’s ability to detect nodes that are critical to maintaining hypergraph connectivity. In addition, we present a detailed parameter analysis of AHGA and compare its performance against several baseline methods. Finally, an ablation study is conducted to highlight the individual contributions of each component within the AHGA framework.
\subsection{Performance evaluation metrics}
\textbf{Kendall's $\tau$ Coefficient~\citep{kendall1938new}:}
Kendall's $\tau$ coefficient is a non-parametric statistic used to evaluate the ordinal association between two ranked variables. In the context of key node identification, it quantifies the global correlation between the predicted ranking scores, denoted as $X = \{x_1, x_2, \ldots, x_N\}$, produced by a given node ranking method, and the ground-truth influence scores $Y = \{y_1, y_2, \ldots, y_N\}$, obtained by simulating the hypergraph-based SIR model with each node individually seeded. A pair of observations $(x_i, y_i)$ and $(x_j, y_j)$ is considered concordant if the orderings of both variables agree, i.e., either $(x_j > x_i \land y_j > y_i)$ or $(x_j < x_i \land y_j < y_i)$; otherwise, the pair is discordant. The Kendall's $\tau$ coefficient is computed as:
\begin{equation}
    \tau=\frac{P - Q}{\frac{1}{2}n(n - 1)},
\end{equation}
where $P$ and $Q$ denote the number of concordant and discordant pairs, respectively, and $n$ is the number of elements being ranked.

\textbf{Rank Overlap~\citep{strona2015new}:}
Kendall’s $\tau$ coefficient quantifies the overall consistency between two node rankings. However, in many real-world scenarios, identifying the top-ranked nodes is of greater practical significance. To assess the effectiveness of our algorithm in capturing these influential nodes, we employ the rank overlap metric, defined as:

\begin{equation}
    \text{O}(T_f,P_f)=\frac{|T_f\cap P_f|}{N_f}\times100\%,
\end{equation}
where $T_f$ represents the top $f\%$ of nodes in the ground-truth ranking derived from the hypergraph-based SIR model, and $P_f$ denotes the top $f\%$ of nodes identified by the evaluated method, $N_f$ is the top $f\%$ number of nodes, and $f \in \{5, 10, 15, 20, 25\}$ is a tunable threshold that determines the evaluation range.

\textbf{$s$-efficiency~\citep{latora2001efficient,xie2023vital}:} In addition to evaluating a node's spreading capability, it is also essential to assess its functional role in facilitating information exchange within a hypergraph. To this end, we adopt the $s$-efficiency metric to quantify the efficiency of communication. The underlying principle of $s$-efficiency is that the effectiveness of information transfer between two nodes diminishes with increasing distance. Formally, the metric is defined as:
\begin{equation}
\mathscr{E}_s(G^H) = \binom{|E_s| }{2}^{-1} \sum_{\substack{e_g, e_q \in E_s \\ g \neq q}} \frac{1}{d_s^e(g, q)},
\end{equation}
where $|E_s|$ denotes the number of hyperedges containing at least $s$ nodes, and $d_s^e(g, q)$ is the $s$-distance between hyperedges $e_g$ and $e_q$. The term $\binom{|E_s|}{2}$ represents the total number of hyperedge pairs under consideration. This metric computes the average inverse $s$-distance across all such hyperedge pairs, providing a measure of how efficiently information can propagate between them. A higher value of $\mathscr{E}_s$ indicates greater overall communication efficiency within the hypergraph’s higher-order structure.

For a given node ranking method, we first sort the nodes based on their computed importance scores. We then remove the top $p$ proportion of nodes from the hypergraph and denote the remaining hypergraph as $G^H_p$. The $s$-efficiency of $G^H_p$ is denoted as $\mathscr{E}_s(G^H_p)$. As $s$-efficiency is highly sensitive to the $s$-distance among hyperedges, it serves as an effective indicator of structural integrity. Based on this, we introduce the following metric to quantify the extent to which the removal of top-ranked nodes degrades the hypergraph’s communication efficiency:
\begin{equation}
\Delta \mathscr{E}(p)=\sum_{s = 1}^{s'} \Delta \mathscr{E}_{s}(p),
\end{equation}
where $\Delta \mathscr{E}_{s}(p) = \mathscr{E}_s(G^H) - \mathscr{E}_s(G^H_p)$ denotes the reduction in $s$-efficiency resulting from the removal of the top $p$ proportion of nodes. A larger value of $\Delta \mathscr{E}(p)$ indicates a greater decline in communication efficiency, reflecting stronger disruptive power of the removed nodes and, consequently, better performance of the ranking method. As previous studies have shown that the trends of $\Delta \mathscr{E}(p)$ are generally consistent across different values of $s$, we set the maximum $s$-distance $s' = 6$ in our experiments as a representative middle value~\citep{xie2023vital}. By aggregating the $s$-efficiency losses across multiple $s$ values, this metric provides a more comprehensive assessment of how the removal of influential nodes affects the overall structural efficiency of the hypergraph.
 
\begin{figure*} 
    \centering
    \includegraphics[width=1.0\linewidth]{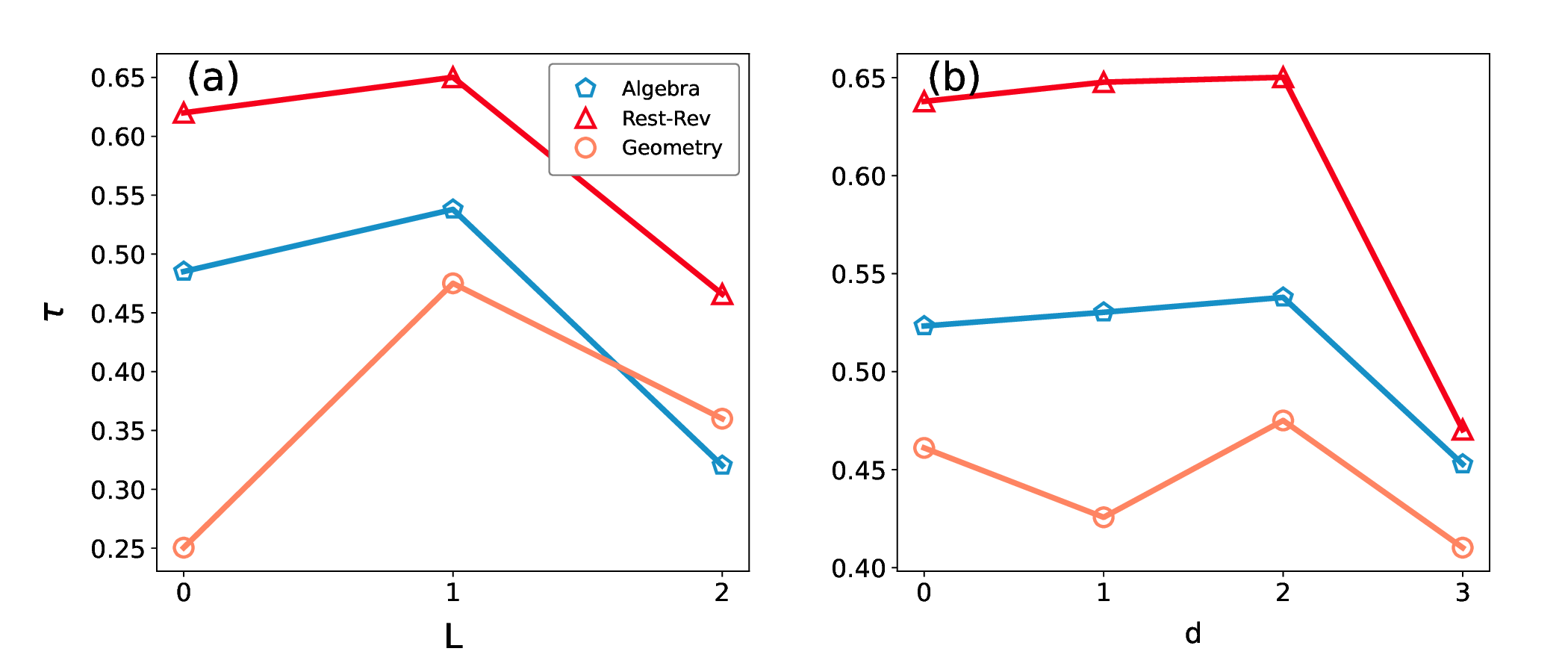}
    \caption{Parameter analysis of AHGA: (a) impact of the number of HGNN layers $L$; (b) impact of the embedding dimension $d$ on model performance.}
    \label{figure4}
\end{figure*}

Subsequently, we analyze the effect of representative nodes identified using different $s$-distances on the Kendall's $\tau$ coefficient, as reported in Table~\ref{table2}. We vary the $s$ value from 1 to 9 with a step size of 1, while excluding $s > 9$ to reduce computational cost. For hypergraphs, e.g., Algebra, Rest-Rev, Geometry, Music-Rev, Email-Enron, and Email-W3C, we observe a decreasing trend in Kendall’s $\tau$ beyond a certain point. As a result, we further omit the evaluation for $s > 6$ on these datasets to avoid unnecessary computation.
The highest $\tau$ values are highlighted in bold. When $s=2$, the hypergraphs Geometry, Email-Enron, and Email-W3C achieve the best performance, with $\tau$ values of 0.4752, 0.6306, and 0.5991, respectively. For $s=3$, Algebra, Rest-Rev, and Music-Rev exhibit peak performance with $\tau$ values of 0.5379, 0.6502, and 0.2276. Notably, Senate-Com and House-Com require higher-order distances to obtain suitable representative nodes, achieving their highest $\tau$ values when $s=9$ and $s=7$, respectively. These observations are consistent with the $s$-distance distributions shown in Figure~\ref{figure4}, where the $s$-distances for Senate-Com and House-Com are generally longer. This indicates that, for such hypergraphs, higher-order distances are necessary to effectively compute node ranking scores. After obtaining the optimal parameter values, we apply them in the subsequent experiments, specifically setting $L=2$, $d=256$, and selecting the optimal $s$ for each hypergraph as indicated in Table~\ref{table2}.

\begin{table*}[t]\footnotesize
\centering
\caption{Parameter analysis of $s$ in the active learning for selecting representative nodes, evaluated by Kendall’s $\tau$. The selected nodes are used to fine-tune the pre-trained model to achieve optimal performance. “Basic” denotes the performance of the pre-trained model without active learning. }
\vspace{8pt}
\renewcommand{\arraystretch}{1}
\setlength{\tabcolsep}{1.2pt}
\begin{tabular}{cccccccccccc}
    \toprule
    {Hypergraphs} & {Basic} & {$s=1$} & {$s=2$}  & {$s=3$} & {$s=4$} & {$s=5$}&{$s=6$}&{$s=7$}&{$s=8$}&{$s=9$} \\
    \midrule
    % \multirow{8}{*}{Empirical} 
     Senate-Com & 0.1613 & 0.1432 & 0.1401 & 0.1326 & 0.1512 & 0.1577& 0.1702 &0.1833& 0.1924& \textbf{0.2431} \\
     Algebra & 0.5015 & 0.4872& 0.5112 &  \textbf{0.5379} & 0.5220 & 0.4721 &0.4892 &$\ast$ &$\ast$ &$\ast$ \\
     Rest-Rev & 0.6291 & 0.6407 & 0.6123 &  \textbf{0.6502} & 0.6210 & 0.5788&0.4872 &$\ast$ &$\ast$ &$\ast$\\
     Geometry & 0.4336 & 0.4230&  \textbf{0.4751} & 0.4622 & 0.4433 & 0.4116 &0.4201 &$\ast$ &$\ast$ &$\ast$\\
     Music-Rev & 0.1724 & 0.1835 & 0.1624 &  \textbf{0.2276} & 0.2031 & 0.2127&0.1653 &$\ast$ &$\ast$ &$\ast$\\
     House-Com & 0.1233 & 0.1054 & 0.0988 & 0.1102 & 0.1379 & 0.1734&0.2079 &\textbf{0.2364}&0.2331&0.2276 \\
     Email-Enron & 0.6201 & 0.6235&  \textbf{0.6306} &0.6297 & 0.6278 & 0.6199 & 0.6071 &$\ast$ &$\ast$ &$\ast$\\
     Email-W3C & 0.5761 & 0.5903 &  \textbf{0.5991} & 0.5706 & 0.5536 & 0.5624 &0.5189 &$\ast$ &$\ast$ &$\ast$\\
    \bottomrule

    \label{table2}
\end{tabular}
\end{table*}

\subsection{Parameter analysis}
In AHGA, key parameters such as the number of HGNN layers $L$, the embedding dimension $d$ in the Autoencoder, and the $s$-distance used in the active learning component can significantly influence the effectiveness of node ranking. To assess the sensitivity of the hyperparameters $L$ and $d$, we conduct parameter tuning experiments on three representative empirical hypergraphs: Algebra, Rest-Rev, and Geometry. The corresponding Kendall's $\tau$ coefficients are presented in Figure~\ref{fig4}.
The parameter $L$ controls the number of HGNN layers and thus determines the range of neighborhood information aggregated. A larger $L$ integrates features from more distant neighbors. As shown in Figure~\ref{fig4}(a), AHGA achieves the best performance with $L=2$ across all three datasets, while overly small or large values of $L$ tend to degrade performance due to insufficient information capture or the introduction of noise.
For the embedding dimension $d$ in the Autoencoder, the performance increases when $d \leq 256$, with the best performance observed at $d=256$. However, increasing $d$ to 512 leads to a notable drop in performance, likely due to overfitting or redundant representations.

\begin{figure}[h] 
     \centering
     \includegraphics[width=.95\linewidth]{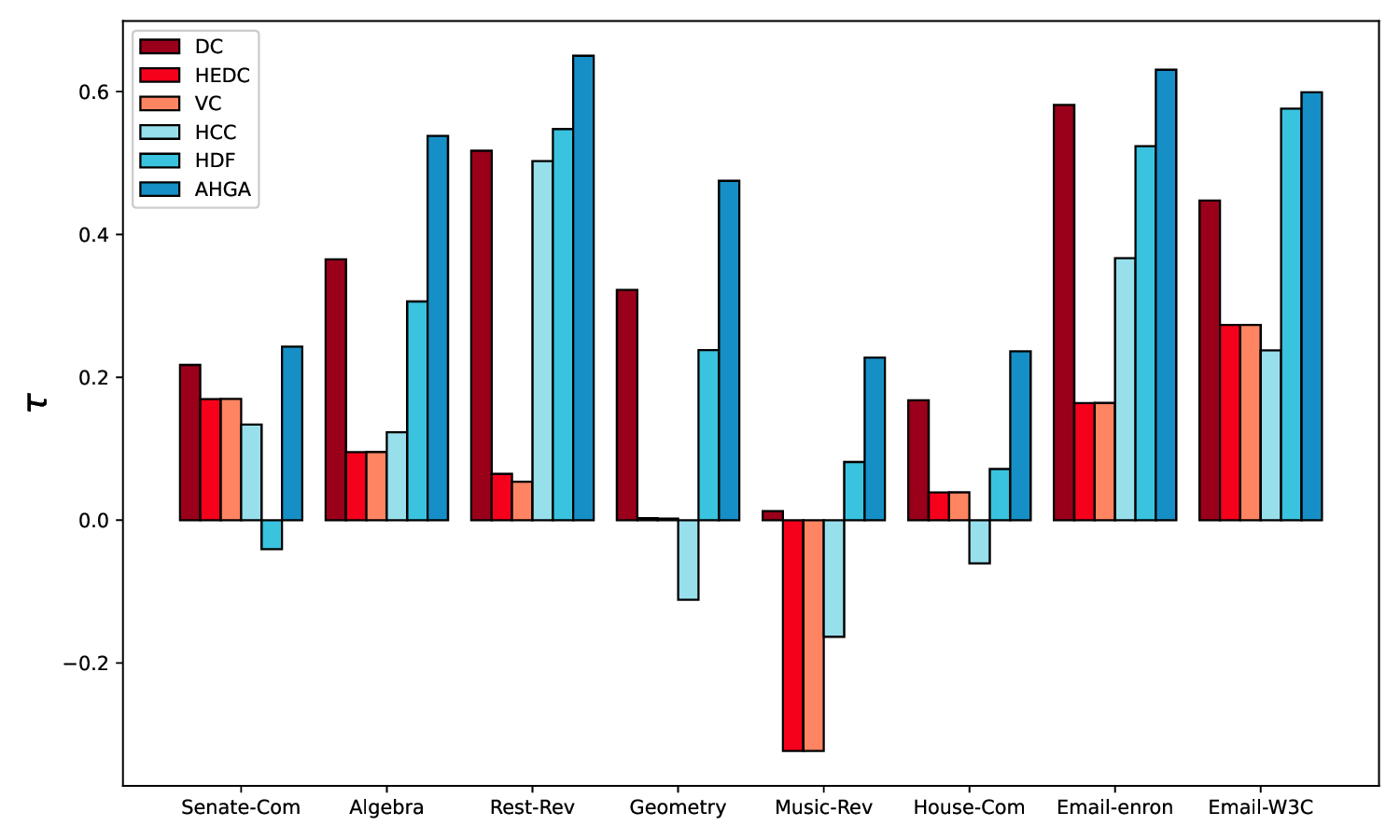}
     \caption{Performance of node ranking methods measured by Kendall's $\tau$ coefficient on empirical hypergraphs.}
     \label{figure5}
 \end{figure}
\subsection{Results}
To assess the effectiveness of the AHGA model, we present the Kendall's $\tau$ coefficients of various methods in Figure~\ref{figure5}. AHGA consistently achieves the highest performance across all empirical hypergraphs, with particularly notable improvements in Algebra, Geometry, Rest-Rev, and Music-Rev. DC also performs robustly, ranking second in most cases. On average, AHGA outperforms DC by 37.4\%. However, we observe relatively low Kendall's $\tau$ values across all methods on Senate-Com, Music-Rev, and House-Com, with some even yielding negative coefficients. Referring to the topological properties in Table~\ref{table1}, these hypergraphs exhibit low coefficients of variation in node degree ($CV(K^V)$), indicating a more homogeneous degree distribution. Such homogeneity reduces the distinguishability of nodes, limiting the effectiveness of most ranking methods. Among hyperedge-based methods, HEDC, VC, and HCC assign equal centrality scores to all nodes within a hyperedge. This uniform allocation overlooks the heterogeneous roles of nodes, resulting in poor performance across all datasets. In contrast, HDF, a method based on higher-order distances, displays unstable results, ranking second only in Rest-Rev and Email-W3C. This variability suggests that the effectiveness of HDF is sensitive to the selection of the order parameter, highlighting the importance of order-specific configurations for hypergraphs with differing structural characteristics.

\begin{figure}[h] 
     \centering
     \includegraphics[width=1\linewidth]{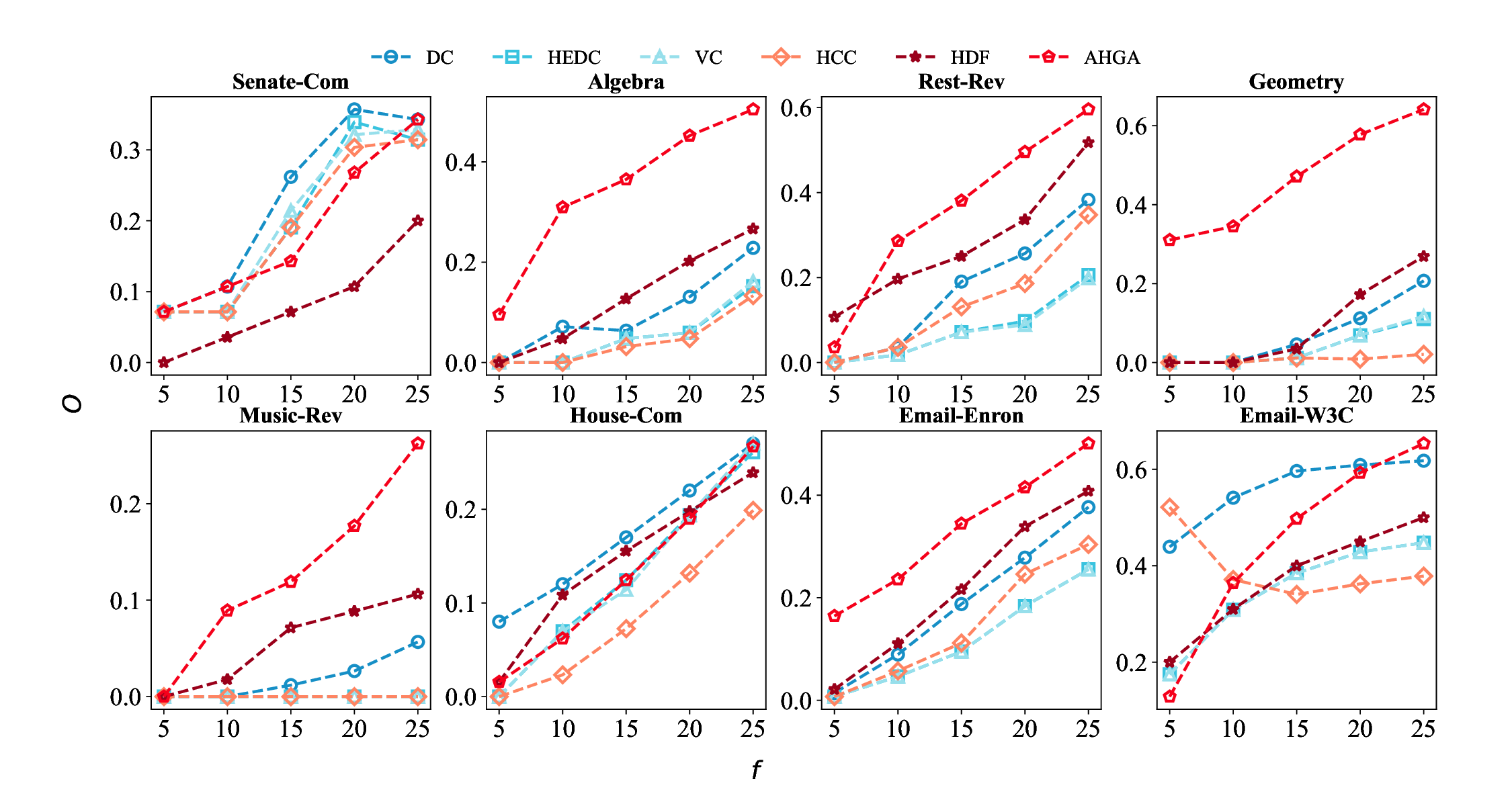}
     \caption{Overlap $O$ between the top $f$ nodes identified by different node ranking methods and the ground-truth influential nodes obtained from Monte Carlo simulations using the hypergraph-based SIR model.}
     \label{figure6}
 \end{figure}

To evaluate the effectiveness of node ranking methods in identifying top influential nodes within a hypergraph, we measure the overlap $O$ between the top-$f$ nodes selected by each method and those obtained via Monte Carlo simulations based on the hypergraph-based SIR model. As previously noted, a higher $O$ value indicates better identification performance. As shown in Figure~\ref{figure6}, AHGA exhibits superior performance, particularly on the Algebra, Geometry, and Rest-Rev hypergraphs. The DC and HDF methods are closely followed, and in certain cases, such as House-Com, Senate-Com, and Email-W3C, they slightly outperform AHGA. An analysis of the structural properties in Table~\ref{table1} reveals that extremely large or small average hyperedge sizes ($\langle K^E \rangle$) tend to generate similar features among certain nodes. This redundancy introduces noise, thereby hindering the predictive accuracy of the models. Nonetheless, as the proportion of selected nodes increases to 25\%, AHGA continues to maintain a distinct advantage.

\begin{figure}[h] 
    \centering
    \includegraphics[width=1\linewidth]{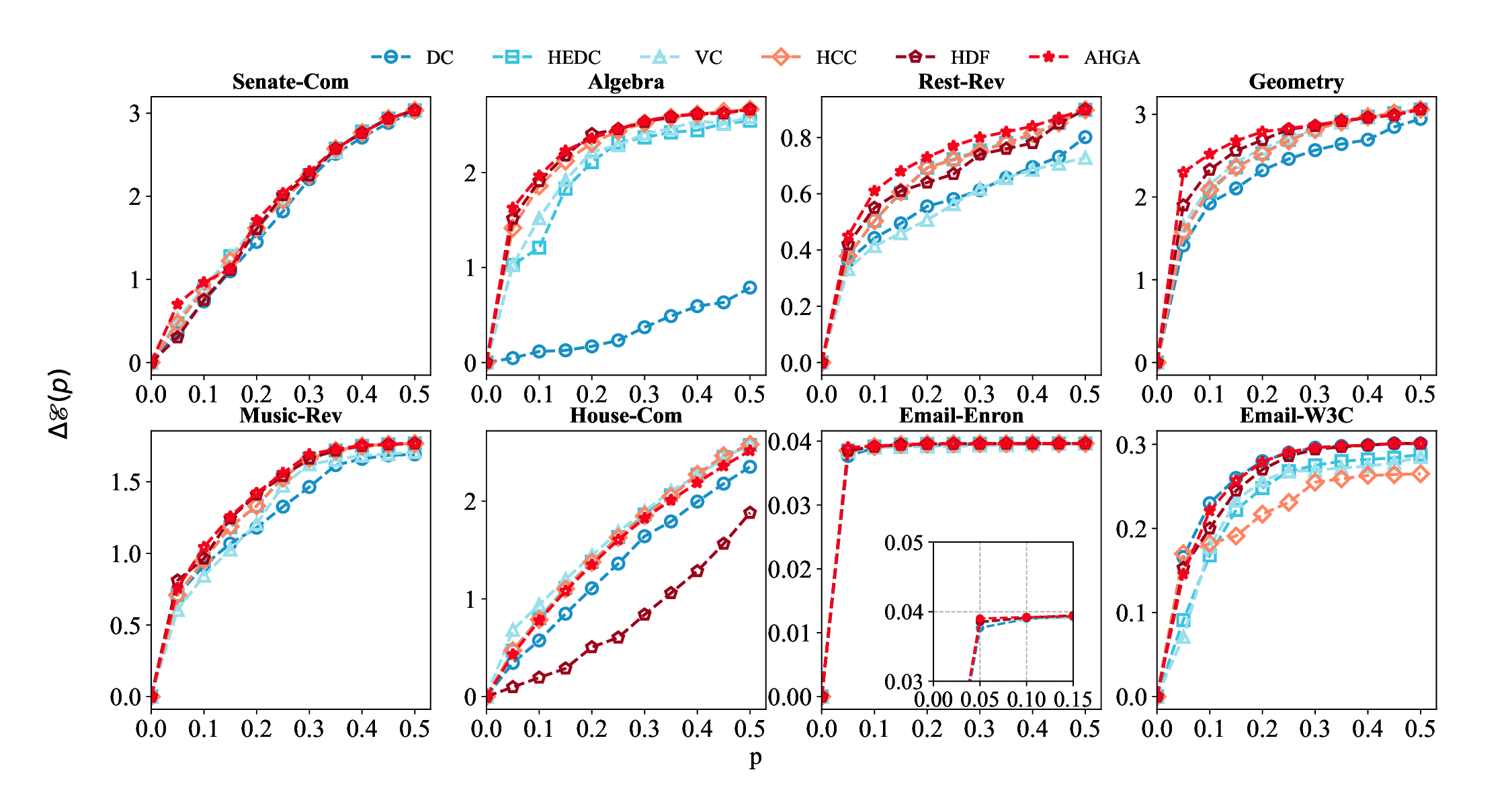}
    \caption{Performance evaluation of node ranking methods in terms of hypergraph connectivity. For each method, we remove a fraction $p$ ($p \in [0, 0.5]$) of top-ranked nodes and compute the corresponding efficiency loss $\Delta \mathscr{E}(p)$, where larger values indicate greater impact on hypergraph structure.}
    \label{figure7}
\end{figure}

We evaluate the performance of the nodes identified by the AHGA model in terms of hypergraph connectivity, as shown in Figure~\ref{figure7}. Specifically, we remove a fraction $p$ ($p \in [0, 0.5]$) of nodes ranked by a specific method and then calculate the corresponding $\Delta \mathscr{E}(p)$. The results reveal that our method outperforms other benchmark approaches on the majority of hypergraphs, with only a slight performance gap observed in the House-Com hypergraph. This suggests that the AHGA model also achieves strong performance in hypergraph dismantling tasks. In contrast, the efficiency of the other centrality-based methods shows considerable variability. Moreover, although DC ranks as the second-best method for identifying influential nodes, it fails to pinpoint those nodes whose removal would lead to significant disruption of the hypergraph structure. Taking Figure \ref{figure6} and Figure \ref{figure7} into account, DC proves relatively effective in identifying highly influential nodes, but these nodes perform poorly in highly coupled hypergraph scenarios such as dismantling. In contrast, AHGA demonstrates greater potential and applicability.

\begin{figure}[] 
    \centering
    \includegraphics[width=0.7\linewidth]{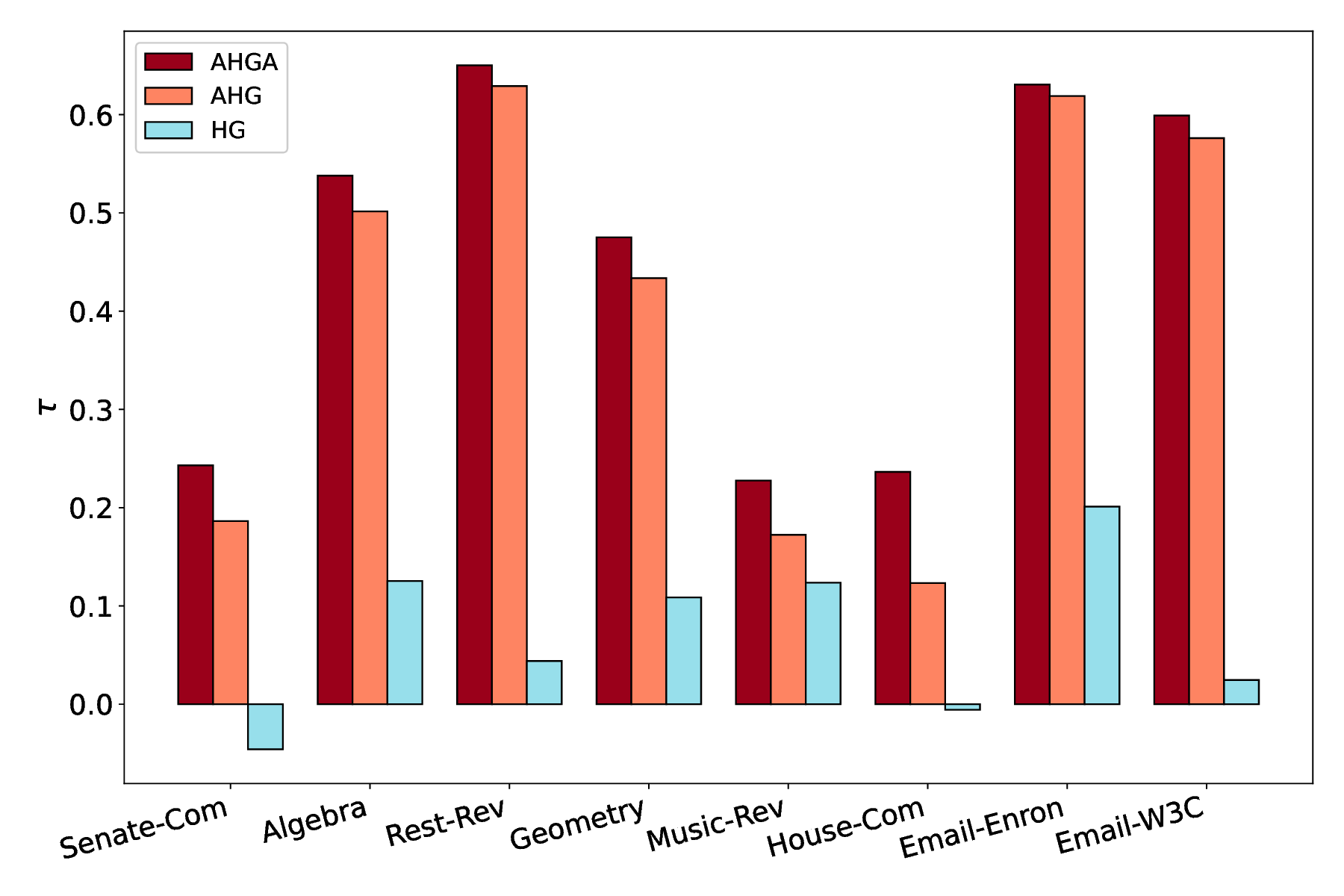}
    \caption{Ablation study on the AHGA model design. AHGA denotes the full model. AHG refers to the variant without the fine-tuning (active learning) module. HG is a further simplified version with both the fine-tuning and feature extraction (autoencoder) modules removed.}
    \label{figure8}
\end{figure}

Within the AHGA framework, we integrate an autoencoder with active learning to identify vital nodes. To evaluate the individual contribution of each component to the overall performance, we perform an ablation study. Specifically, we report the Kendall’s $\tau$ values for two variants: AHG, which excludes the active learning module, and HG, which omits both the autoencoder and active learning. As illustrated in Figure~\ref{figure8}, the autoencoder contributes more substantially to the model’s performance, while the addition of the fine-tuning process via active learning further improves the results. These findings suggest that the autoencoder effectively encodes node features related to spreading potential, providing a solid foundation for distinguishing node importance. Simultaneously, active learning enhances the model’s adaptability by mitigating the structural discrepancies between training data and real-world hypergraphs.

% \subsection{Practical significance}\label{Practical significance}

% As interactions among entities in many real-world systems are often of a higher order, aligning with the attributes of hyperedges, the method proposed in this paper has broad applications. For example, in social platform governance, it can accurately identify influential bloggers to achieve precise intervention and quickly curb the spread of rumors. In epidemic control, it can identify superspreaders in group contacts, optimize resource allocation, and effectively contain disease transmission through targeted quarantine measures.
% \color{red}}

\section{Conclusions}\label{Conclusions}
The higher-order characteristics intrinsic to hypergraph structures pose new challenges for key node identification. Traditional centrality-based methods often suffer from limited scalability, while existing deep learning approaches tailored for conventional networks fail to effectively capture the complex relational dependencies among nodes in hypergraphs. To address these limitations, we propose the AHGA framework, i.e., a hypergraph neural network-based method, for identifying critical nodes in real-world hypergraphs. AHGA integrates an autoencoder architecture with a pre-training stage and an active learning strategy to enhance both accuracy and efficiency in node ranking.

The autoencoder module is designed to extract higher-order topological features from complex hypergraph structures. Specifically, it reprocesses the global structural information encoded via one-hot representations and local degree-based metrics to generate optimal feature embeddings that accurately characterize each node. As shown in Figure~\ref{figure8}, the autoencoder plays a critical role in enhancing the overall performance of the AHGA model. During the pre-training phase, we construct a three-layer Hypergraph Neural Network (HGNN), which aggregates features from up to three-hop neighbors through a message-passing mechanism. This design enables effective information integration while mitigating the potential noise introduced by incorporating more distant neighbors. Subsequently, the model is fine-tuned via an active learning strategy, which addresses the structural discrepancies between synthetic training data and real-world hypergraphs, thereby improving the robustness and generalizability of AHGA across diverse hypergraph scenarios. Extensive experiments demonstrate that the proposed AHGA model consistently outperforms existing baselines in identifying multifunctional nodes, i.e., those that are not only highly influential but also possess strong hypergraph-disrupting capability.

In light of the emerging nature of critical node identification in hypergraphs, future work will focus on incorporating more efficient and scalable techniques into the AHGA framework to further improve its performance. In particular, the integration of large language models (LLMs) into hypergraph analysis presents a promising avenue for exploration~\citep{mao2024identify}. By leveraging the powerful reasoning and abstraction capabilities of LLMs, it may be possible to extract richer semantic and structural information from complex higher-order interactions and apply this knowledge to enhance node ranking and other downstream tasks. This direction offers considerable potential for advancing the effectiveness and interpretability of key node identification in hypergraphs.
\section*{CRediT authorship contribution statement}

\textbf{Xiaonan Ni:} Writing - original draft, Writing -- review \& editing, Visualization, Validation, Supervision, Software, Project administration, Methodology, Investigation, Conceptualization. \textbf{Guangyuan Mei:} Writing -- review \& editing, Validation, Writing - original draft, Project administration. \textbf{Su-Su Zhang:} Writing -- review \& editing, Validation, Project administration. \textbf{Yang Chen:}  Writing – review \& editing, Supervision. \textbf{Xin Xu:} 
Writing -- review \& editing, Supervision. \textbf{Chuang Liu:} Writing -- review \& editing, Supervision.   \textbf{Xiu-Xiu Zhan}: Writing -- review \& editing, Validation, Supervision, Project administration.

\section*{Declaration of competing interest}

The authors declare that they have no known competing financial interests or personal relationships that could have appeared to influence the work reported in this paper.

\section*{Acknowledgements}

This work was supported by the China Postdoctoral Science Foundation (2024M762809), and the National Natural Science Foundation of China (Grant No. 62473123).

% \section*{Author Contributions Statement}
% \
% All authors planed the study and the experiment. All authors conducted the calculations and simulations. All authors developed the analytical tools and analyzed the results. All authors wrote and approved the final manuscript.

% \section*{Data and Code Availability}
% \begin{sloppypar}
% The data and codes used for this study are available at: https://github.com/Slot-Ni/AHGA.git.
% \end{sloppypar}

\clearpage

\bibliography{TempExample.bib}

% \section*{Acknowledgements (not compulsory)}

% Acknowledgements should be brief, and should not include thanks to anonymous referees and editors, or effusive comments. Grant or contribution numbers may be acknowledged.

% \section*{Author contributions statement}

% Must include all authors, identified by initials, for example:
% A.A. conceived the experiment(s),  A.A. and B.A. conducted the experiment(s), C.A. and D.A. analysed the results.  All authors reviewed the manuscript.

\end{document}